\documentclass[twocolumn,prl]{revtex4}
\usepackage[dvips]{graphicx,color}
\usepackage{graphicx}
\usepackage{amsmath}
\usepackage{amssymb}
\usepackage{bm}
\usepackage{latexsym}
\newcommand{\op}[1]{\hat {#1}}

\def\vect#1{\text{\boldmath $#1$}}

\newenvironment{SecIntro}{}{}
\newenvironment{SecTheory}{}{}
\newenvironment{SecExSetup}{}{}
\newenvironment{SecExResults}{}{}
\newenvironment{SecConclusion}{}{}
\newenvironment{spmatrix}{\left(\begin{smallmatrix}}{\end{smallmatrix}\right)}

\begin{document}

\title{Demonstration of a Controlled-Phase Gate \\ for Continuous-Variable One-Way Quantum Computation}

\author
{Ryuji Ukai$^{1}$, Shota Yokoyama$^{1}$, Jun-ichi Yoshikawa$^{1}$, Peter van Loock$^{2,3}$, and Akira Furusawa$^{1}$}

\affiliation{$^{1}$Department of Applied Physics, School of Engineering, The University of Tokyo,\\ 7-3-1 Hongo, Bunkyo-ku, Tokyo 113-8656, Japan\\
$^{2}$Optical Quantum Information Theory Group, Max Planck Institute for the Science of Light, G\"unther-Scharowsky-Str.1/Bau 26, 91058 Erlangen, Germany\\
$^{3}$Institute of Theoretical Physics I, Universit\"at Erlangen-N\"urnberg, Staudstr.7/B2, 91058 Erlangen, Germany}

\begin{abstract}
We experimentally demonstrate a controlled-phase gate for continuous variables in a fully
measurement-based fashion. In our scheme,
the two independent input states of the gate, encoded in two optical modes,
are teleported into a four-mode Gaussian cluster state. As a result, one of the entanglement links
present in the initial cluster state appears in the two unmeasured output modes
as the corresponding entangling gate acting on the input states.
The genuine quantum character of this gate becomes manifest and is verified through the presence of entanglement
at the output for a product two-mode coherent input state.
By combining our controlled-phase gate with the recently reported module for universal single-mode Gaussian operations
[R.\ Ukai {\it et al.}, \prl {\bf 106}, 240504 (2011)], it is possible to implement universal Gaussian operations
on arbitrary multi-mode quantum optical states in form of a fully measurement-based one-way quantum computation.
\end{abstract}

\maketitle

\begin{figure}
\centering
\includegraphics[height=10cm,clip]{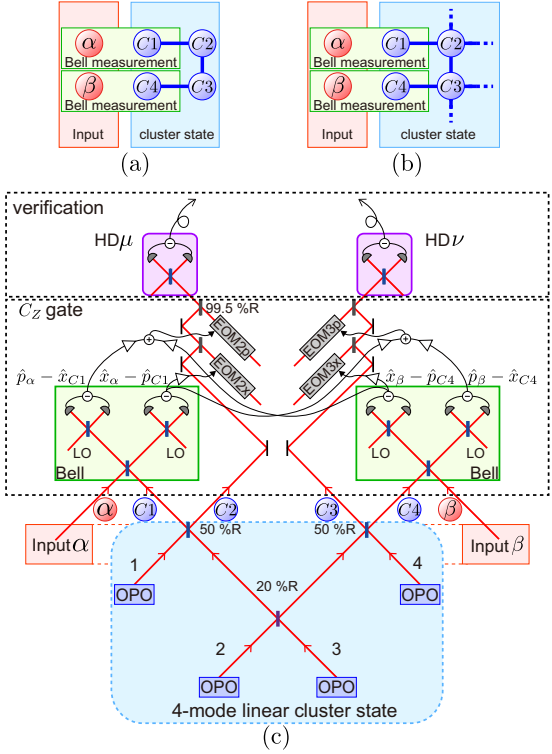}
\caption{(color online) (a) An abstract illustration of our experiment.
(b) Input coupling through quantum teleportation for larger one-way quantum computations.
(c) A schematic of our experimental setup.
OPO: optical parametric oscillator,
LO: local oscillator for homodyne measurement, EOM: electro-optical modulator,
HD: homodyne detection, Bell: Bell measurement.
}
\label{FigAbstractIllustation}
\end{figure}

\begin{SecIntro}
The one-way model of measurement-based quantum computation (QC) \cite{Raussendorf01} is a fascinating alternative
to the standard unitary-gate-based circuit model,
for discrete-variable systems such as qubits as well as for continuous-variable (CV) encodings
on quantized harmonic oscillators
\cite{Zhang06,Menicucci06,Furusawa11}.
Such one-way computations are realized through single-qubit or single-mode measurements,
and some outcome-dependent feedforward operations,
on a pre-prepared multi-party entangled state, the so-called ``cluster state''.
By choosing an appropriate set of measurement bases on a sufficiently large cluster state,
an arbitrary unitary operation can be implemented for the corresponding encoding.

Towards CV QC,
initially, a CV analogue to qubit cluster states was proposed \cite{Zhang06}.
Subsequently, the notion of an in-principle universality of CV one-way quantum computation was then proven,
relying upon the asymptotic assumptions of sufficiently long measurement-based gate
sequences \cite{Lloyd99} and perfectly squeezed optical cluster-state resources
as well as the inclusion of at least one nonlinear, non-Gaussian measurement device \cite{Menicucci06}.
Only shortly thereafter,
by using squeezed vacuum states and beam splitters \cite{Peter07},
various cluster states were generated in the lab \cite{Su07,Yukawa08C}.
Among these was the four-mode linear cluster state which would
allow to implement arbitrary single-mode Gaussian operations \cite{Peter07J,Ukai09}.
However, in order to demonstrate such single-mode gate operations on
arbitrary input states, 
the input mode has to be attached to the cluster state.
For a single quadratic gate, this can be accomplished 
using two squeezed-state ancillae, as described in Ref.~\cite{Miwa09}.
A much simpler and more general solution \cite{Ukai09}, however, would employ
a multi-mode measurement such as a Bell measurement, similar 
to standard CV quantum teleportation \cite{Furusawa98}.
By using a four-mode linear cluster state and a Bell-measurement-based 
input coupling,
a typical set of single-mode Gaussian operations
such as squeezing and Fourier transformations was recently experimentally demonstrated
\cite{UkaiExOne10}.

The final missing element towards implementing arbitrary 
multi-mode Gaussian transformations in a one-way fashion \cite{Ukai09}
is a universal two-mode entangling gate.
In fact, universal multi-mode operations (even including non-Gaussian ones)
are, in an asymptotic sense, realizable when universal single-mode gates 
(including at least one non-Gaussian gate) are combined
with any kind of quadratic (Gaussian) interaction gate \cite{Lloyd99}.
More specifically, an arbitrary multi-mode Gaussian operation
can be exactly recast as a finite decomposition into single-mode Gaussian gates
and a quadratic (Gaussian) two-mode gate \cite{Lloyd99,Braunstein2005,Reck,Ukai09}.
The most natural and easily implementable two-mode gate
in this setting would be that corresponding to a vertical link between
two individual modes of a CV cluster state -- the controlled-phase ($C_Z$) gate
\cite{Menicucci06}.
Very recently, Wang \textit{et al.} reported an attempt to experimentally demonstrate this gate \cite{Wang10}.
However, the cluster state in that experiment was not of sufficient quality in order
to operate the gate as a genuine nonclassical entangling gate; in fact,
the two-mode input quantum state was degraded by such a large excess noise that
no entanglement at all was present at the output.

In this paper, we demonstrate a CV cluster-based $C_Z$ gate operating in the quantum realm.
In order to verify its nonclassicality, we show that the two-mode output state is indeed entangled
when the input state is a product of two single-mode coherent states. Furthermore,
several manifestations of
the general input-output relation of the gate are realized, 
using various distinct coherent states as the input of the gate. 
The resource state is a four-mode linear cluster state,
as illustrated in Fig.~\ref{FigAbstractIllustation}(a).
The two input coherent states are prepared independently of the cluster state and subsequently coupled to it
through quantum teleportations.
Note that a similar experiment in the qubit regime was reported recently \cite{WeiBo10PNAS}.
Our CV gate can be directly incorporated into a one-way quantum computation of larger scale,
see Fig.~\ref{FigAbstractIllustation}(b) \cite{LargeOneWay}.
In particular, as noted above, when combined with universal single-mode gates,
arbitrary multi-mode operations are in principle realizable; and certain nonlinear 
multi-mode Hamiltonians such as a fairly strong two-mode cross-Kerr interaction (a universal entangling gate
for photonic qubits \cite{ChuangYamamoto} when effectively enhanced) would only require applying tens of quadratic and cubic 
single-mode gates, in addition to the two-mode $C_Z$ gate \cite{SefiVanLoock}.

\end{SecIntro}

\begin{figure*}
\centering
\includegraphics[height=3.8cm,clip]{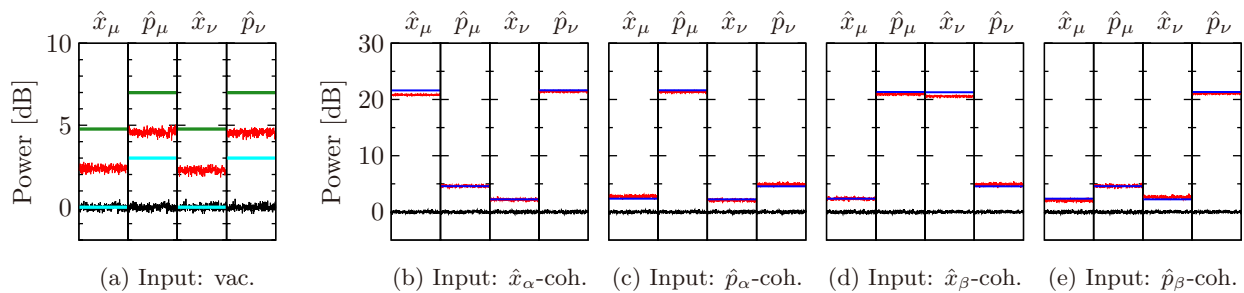}
\caption{(color)
Powers at the outputs.
(a) shows variances of the output quadratures for vacuum inputs.
The black and red traces correspond to the shot noise levels (SNLs) and output quadratures, respectively.
The green lines show the theoretical prediction when no resource squeezing is available, while
the cyan traces show the theoretical prediction for an ideal $C_Z$ gate.
(b), (c), (d), and (e) show the powers of the output quadratures where
($\left<\op{x}_\alpha\right>$,$\left<\op{p}_\alpha\right>$,$\left<\op{x}_\beta\right>$,$\left<\op{p}_\beta\right>$) are
($a$,0,0,0), (0,$a$,0,0), (0,0,$b$,0), and (0,0,0,$b$),
and where $a$ and $b$ correspond to 21.5 dB and 21.2 dB above the SNL, respectively.
The blue lines show the theoretical prediction based on (a) and
different input coherent amplitudes.
Vac.: vacuum state.
Coh.: coherent state.
}
\label{FigCoherentResults}
\end{figure*}

\begin{SecTheory}

In the following, we shall use the canonical 
position and momentum operators, $\op{x}_j$ and $\op{p}_j$,
where the subscript $j$ denotes an optical mode and $[\op{x}_j,\op{p}_k]=i\delta_{jk}/2$.
The CV $C_Z$ gate corresponds to the unitary  operator $\op{C}_{Zjk}=e^{2i\op{x}_j\op{x}_k}$
with the input-output relation,
$
\op{\vect{\xi}}'_{jk}=\begin{spmatrix}
I & S \\
S & I
\end{spmatrix}
\op{\vect{\xi}}_{jk}
$,
where $\op{\vect{\xi}}_{jk} = (\op{x}_j,\op{p}_j,\op{x}_k,\op{p}_k)^T$,
$S=\begin{spmatrix}
0 & 0 \\
1 & 0
\end{spmatrix}$,
and $I$ is the $2\times 2$ identity matrix.

We demonstrate this $C_Z$ gate by using a four-mode linear cluster state [$C1$-$C2$-$C3$-$C4$ in Fig.~\ref{FigAbstractIllustation}(a)].
A CV cluster state is defined, in the ideal case, through
its zero eigenvalues for certain linear combinations of the canonical operators,
$\op{p}_{Cj}-\sum_{k\in {N_j}}\op{x}_{Ck}(\equiv\op{\delta}_j)$.
Here, $N_j$ denotes the set of nearest-neighbor modes of mode $j$,
when the state is represented by a graph, see Figs.~\ref{FigAbstractIllustation}(a-b).
The four-mode linear cluster state can be interpreted
as two Einstein-Podolsky-Rosen (EPR) pairs ($C1$-$C2$ and $C3$-$C4$) with a $C_Z$ interaction
between them ($C2$-$C3$), up to local phase rotations.
When two input states encoded in modes $\alpha$ and $\beta$ are teleported
to modes $C2$ and $C3$, using the double instance of EPR states,
the initial $C_Z$ interaction between the two EPR pairs is teleported onto
the two input states \cite{LargeOneWay}.

Let us describe the above procedure for a non-ideal, finitely squeezed cluster state
corresponding to non-zero variances for the operators $\op{\delta}_j$.
When a four-mode linear cluster state is generated by using four finitely squeezed states
and three beam splitters as in Fig.~\ref{FigAbstractIllustation}(c),
the excess noises are as follows,
$\op{\delta}_1=\sqrt{2}e^{-r}\op{p}^{(0)}_1$,
$\op{\delta}_2=-\frac{5}{\sqrt{10}}e^{-r}\op{p}^{(0)}_3-\frac{1}{\sqrt{2}}e^{-r}\op{p}^{(0)}_4$,
$\op{\delta}_3=\frac{1}{\sqrt{2}}e^{-r}\op{p}^{(0)}_1-\frac{5}{\sqrt{10}}e^{-r}\op{p}^{(0)}_2$, and
$\op{\delta}_4=-\sqrt{2}e^{-r}\op{p}^{(0)}_4$,
where $e^{-r}\op{p}_j^{(0)}$ is the squeezed quadrature of the $j$-th squeezed state.
Here, we assume identical squeezing levels with parameter $r$ for simplicity.
Note that the limit $r\to\infty$ corresponds to an ideal cluster state.
Two pairs of modes, ($\alpha,C1$) and ($\beta,C4$), are then subject to Bell measurements.
For this purpose, four observables,
$
\op{p}_\alpha-\op{x}_{C1}, \op{x}_\alpha-\op{p}_{C1},
\op{p}_\beta-\op{x}_{C4},
$
and
$\op{x}_\beta-\op{p}_{C4}$
are measured,
giving the measurement results $t_\alpha$, $t_1$, $t_\beta$, and $t_4$, respectively.
Then the corresponding feedforward operations
$\op{X}_{C2}(t_1)\op{Z}_{C2}(t_\alpha+t_4)\op{X}_{C3}(t_4)\op{Z}_{C3}(t_\beta+t_1)$
are performed on modes $C2$ and $C3$,
where $\op{X}_j(s)=e^{-2is\op{p}_j}$ and $\op{Z}_j(s)=e^{2is\op{x}_j}$
are the position and momentum displacement operators.
The resulting position and momentum operators of modes $C2$ and $C3$ at the output,
labeled by $\mu$ and $\nu$, can then be written as,
\begin{align}
\op{\vect{\xi}}_{\mu\nu}=
\begin{pmatrix}
I & S \\
S & I
\end{pmatrix}
\op{\vect{\xi}}_{\alpha\beta}
+\op{\vect{\delta}}.
\label{EqInOutOfCZ}
\end{align}
This completes the $C_Z$ operation.
Here,
$\op{\vect{\delta}}=(-\op{\delta}_1,
-\op{\delta}_4+\op{\delta}_2 ,
-\op{\delta}_4 ,
-\op{\delta}_1 +\op{\delta}_3)^T$
represents the excess noise of our $C_Z$ gate.
In the ideal case with $r\to \infty$, the noise term $\op{\vect{\delta}}$
vanishes and a perfect $C_Z$ operation is achieved.
As the $C_Z$ gate is an entangling gate,
the presence of entanglement at the output for a product input state
(in spite of the excess noise $\op{\vect{\delta}}$) is crucial to
prove the nonclassicality of our gate implementation.
A sufficient condition for inseparability of a two-mode state is
$
\left<\Delta^2(g\op{p}_\mu-\op{x}_\nu)\right> + \left<\Delta^2(g\op{p}_\nu-\op{x}_\mu)\right> < g
$ for some $g\in \mathbb{R}$ \cite{Duan00,Peter03,Yoshikawa08}.
We will show that this inequality is satisfied at the output for a two-mode vacuum input.
In our case, $g=3/4$ gives the minimal resource requirement,
$
e^{-2r}  <2/5
$,
corresponding to approximately $-4.0$ dB squeezing.

\end{SecTheory}

\begin{figure}
\centering
\includegraphics[height=8cm,clip]{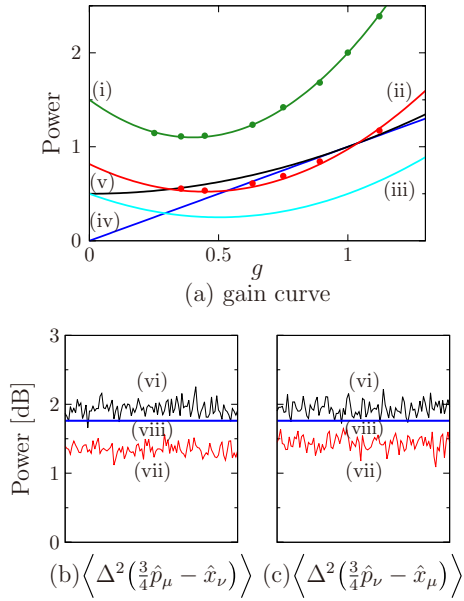}
\caption{(color online)
Entanglement at the output.
(a) shows entanglement verification with several gains $g$.
(i) without resource squeezing,
(ii) with $-5$ dB resource squeezing,
(iii) with infinite resource squeezing (the ideal case),
(iv) sufficient condition for entangling,
and (v) the reference variance when two homodyne inputs are vacuum states.
(b) and (c) show the variances of
$\left<\Delta^2(g\op{p}_\mu-\op{x}_\nu)\right>$ and $\left<\Delta^2(g\op{p}_\nu-\op{x}_\mu)\right>$
at $g=3/4$, respectively.
0 dB corresponds to the SNL.
(vi) the reference where two homodyne inputs are vacuum states,
(vii) the measurement results,
and (viii) sufficient condition for entanglement.
}
\label{FigEntanglingResults}
\end{figure}

\begin{SecExSetup}

A schematic of our experimental setup is illustrated in Fig.~\ref{FigAbstractIllustation}(c).
The light source is a continuous-wave Ti:sapphire laser with a wavelength of 860 nm and a power of 1.7 W.
Four squeezed vacuum states are generated by four optical parametric oscillators (OPOs).
We employ the experimental techniques described in Refs.~\cite{Yukawa08C} and \cite{Yukawa08T}
for the generation of the cluster state and the feedforward process, respectively.
The resource squeezing is $-$5 dB on average and
the detectors' quantum efficiencies are greater than 99\%.
The interference visibilities are 97\% on average, while
the propagation losses from the OPOs to the homodyne detectors are 3-10\%.

\end{SecExSetup}
\begin{SecExResults}

In the following, we show our experimental results.
In Fig.~\ref{FigCoherentResults},
the input-output relation of our gate is investigated,
using several coherent states for the input.
In Fig.~\ref{FigEntanglingResults},
the correlations at the output are determined,
from which the presence of entanglement is verified.
We use a spectrum analyzer to measure the power of the output quadratures.
The measurement frequency is 1 MHz.
The resolution and video bandwidths are 30 kHz and 300 Hz, respectively.
All data in Fig.~\ref{FigCoherentResults} are averaged 20 times,
while those in Fig.~\ref{FigEntanglingResults} are averaged 40 times.

First, Fig.~\ref{FigCoherentResults}(a) shows the variances of the output quadratures
when the two inputs are each in a vacuum state.
In the case of the ideal $C_Z$ gate,
the variances of $\op{x}_\mu$ and $\op{x}_\nu$ remain unchanged and thus are equal to the shot noise level (SNL),
while those of $\op{p}_\mu$ and $\op{p}_\nu$ are 3 dB above the SNL
(two times the SNL) as shown by the cyan lines.
When the resource squeezing is finite,
the output states are degraded by excess noise.
We show as a reference the theoretical prediction
for a vacuum resource [$r=0$ in Eq. \eqref{EqInOutOfCZ}] by green lines,
where the variances of $\op{x}_\mu$ and $\op{x}_\nu$ are
4.8 dB above the SNL (three times the SNL), while those of $\op{p}_\mu$ and $\op{p}_\nu$
7.0 dB above the SNL (five times the SNL).
The experimental results of
$\left<\Delta^2\op{x}_\mu\right>$,
$\left<\Delta^2\op{p}_\mu\right>$,
$\left<\Delta^2\op{x}_\nu\right>$, and
$\left<\Delta^2\op{p}_\nu\right>$,
shown by the red traces, are below the green lines
due to the finite resource squeezing.
These correspond to
2.4 dB,
4.6 dB,
2.2 dB, and
4.6 dB
above the SNL, respectively.
These results are consistent with the resource squeezing level of $-5$ dB,
which leads to 2.1 dB for $\op{x}_\mu$, $\op{x}_\nu$ and
4.7 dB for $\op{p}_\mu$, $\op{p}_\nu$ above the SNL.

In order to verify the general input-output relations,
we employ coherent input states [Figs.~\ref{FigCoherentResults}(b-e)].
The powers of the amplitude quadratures
are measured in advance, corresponding to 21.5 dB for mode $\alpha$
and 21.2 dB for mode $\beta$, respectively, compared to the SNL.

Fig.~\ref{FigCoherentResults}(b) shows the powers of the output quadratures as red traces
when the input $\alpha$ is in a coherent state with a nonzero amplitude only in the $\op{x}_\alpha$ quadrature;
the input $\beta$ is in a vacuum state.
We observe an increase in powers of $\op{x}_\mu$ and $\op{p}_\nu$ caused by
the nonzero coherent amplitude.
On the other hand, $\op{p}_\mu$ and $\op{x}_\nu$ are not changed compared to the case of two vacuum inputs.
In the same figure, the theoretical prediction is shown by blue lines.
Clearly, the experimental results are in agreement with the theory.
Similarly, Figs.~\ref{FigCoherentResults}(c-e) show the results
with a nonzero coherent amplitude in the $\op{p}_\mu$, $\op{x}_\nu$, and $\op{p}_\nu$ quadratures, respectively.
We see the expected feature of the $C_Z$ gate
that the quadratures in modes $\alpha$ and $\beta$ are transmitted to modes $\mu$ and $\nu$ with unity gain
and
$\op{x}_\alpha$ and $\op{x}_\beta$ are transferred to $\op{p}_\nu$ and $\op{p}_\mu$.
We believe that the small discrepancies between our experimental results and the theoretical predictions
are caused by propagation losses and imperfect visibilities.

For assessing the entanglement at the output,
we use two input states in the vacuum.
The two homodyne signals are added electronically with a ratio of
$g^2:1$ and $1:g^2$ in power, by which
$\left<\Delta^2(g\op{p}_\mu-\op{x}_\nu)\right>$ and
$\left<\Delta^2(g\op{p}_\nu-\op{x}_\mu)\right>$ are measured.

Fig.~\ref{FigEntanglingResults}(a) shows the theoretical and experimental results for
$\left<\Delta^2(g\op{p}_\mu-\op{x}_\nu)\right>+\left<\Delta^2(g\op{p}_\nu-\op{x}_\mu)\right>$
with several gains $g$.
The sufficient condition for entanglement is that
$\left<\Delta^2(g\op{p}_\mu-\op{x}_\nu)\right> + \left<\Delta^2(g\op{p}_\nu-\op{x}_\mu)\right>$
is less than $g$, shown by line (iv), for some $g\in \mathbb{R}$.
When $g=0.63$, $0.75$, and $0.89$, this criterion is satisfied in the experiment.
The results without and with resource squeezing
roughly coincide with the theoretical curves (i) and (ii), respectively.
In particular, Figs.~\ref{FigEntanglingResults}(b-c) show the results for
the optimal gain $g=3/4$.
Traces (vi) show the reference for normalization when the two homodyne inputs are vacuum states.
These levels correspond to $1+g^2$ times the SNL.
Traces (vii) show the measurement results for
$\left<\Delta^2(g\op{p}_\mu-\op{x}_\nu)\right>$ and
$\left<\Delta^2(g\op{p}_\nu-\op{x}_\mu)\right>$,
which are $-0.59\pm 0.02$ dB and $-0.50\pm 0.02$ dB
relative to traces (vi), respectively.
Note that the error in determining the SNL is included in the above errors.
Lines (viii) show the sufficient condition for entanglement,
corresponding to the theoretical prediction with about $-4.0$ dB resource squeezing.
The fact that traces (vii) are below lines (viii) proves
that the output state is entangled.
By normalizing the entanglement criterion,
the obtained entanglement is quantified as follows,
\begin{align}
&\left<\Delta^2(\sqrt{g}\op{p}_\mu - \tfrac{1}{\sqrt{g}}\op{x}_\nu)\right>
 + \left<\Delta^2 (\sqrt{g}\op{p}_\nu-\tfrac{1}{\sqrt{g}}\op{x}_\mu)\right>\nonumber\\
&\hspace{7em}= 0.919 \pm 0.003 < 1,\quad \mathrm{at}\,\, g=3/4.
\end{align}

Note that traces (vi), (vii), and line (viii) correspond to
curves (v), (ii), and line (iv) at $g=3/4$, respectively.

\end{SecExResults}

\begin{SecConclusion}
In conclusion, we have experimentally demonstrated
a fully cluster-based $C_Z$ gate for continuous variables.
In our scheme, the two-mode input state was coupled to a four-mode
resource cluster state through quantum teleportations.
For a product input state,
entanglement at the output was clearly observed, verifying the essential property of the $C_Z$ gate.
In combination with our recent work on the experimental demonstration of single-mode Gaussian operations,
all components for universal multi-mode Gaussian operations are now available in a one-way configuration.
The quality of our $C_Z$ gate is only limited by the squeezing level of the resource state,
and the recently reported, higher levels of squeezing \cite{Takeno07,Mehmet09} would even allow
to realize multi-step multi-mode one-way quantum computations. To achieve full universality when processing
arbitrary multi-mode quantum optical states,
the only missing ingredient is a single-mode non-Gaussian gate.

This work was partly supported by SCF, GIA, G-COE, APSA, and FIRST commissioned by the MEXT of Japan,
ASCR-JSPS, and SCOPE program of the MIC of Japan.
R.U. acknowledges support from JSPS.
P.v.L. acknowledges support from the Emmy Noether programme of the DFG in Germany.
\end{SecConclusion}


\begin{thebibliography}{00}
\bibitem{LargeOneWay}
	For deriving this fact, we must only consider the commutation relation
	between the $C_Z$ interaction and the feedforward displacements in the teleportations.
\end{thebibliography}
\end{document}